\documentclass{article}
\usepackage{spconf,amsmath,graphicx,amssymb}
\usepackage{color,multirow}
\usepackage{cite}
\usepackage{type1cm}
\usepackage{setspace}   

\def\etal{\textit{et al. }}

\DeclareMathOperator\SDR{SDR}

\DeclareMathOperator\STPR{STPR}

\DeclareMathOperator\SDRi{SDRi}

\DeclareMathOperator\DISDR{DI_{SDR}}

\title{Amicable examples for informed source separation}
\name{
Naoya Takahashi, Yuki Mitsufuji
}
\address{Sony Group Corporation, Japan}
\begin{document}
\fontsize{9.3pt}{11.4pt}\selectfont


\maketitle
\begin{abstract}
This paper deals with the problem of informed source separation (ISS), where the sources are accessible during the so-called \textit{encoding} stage. Previous works computed side-information during the encoding stage and source separation models were designed to utilize the side-information to improve the separation performance. 
In contrast, in this work, we improve the performance of a pretrained separation model that does not use any side-information. To this end, we propose to adopt an adversarial attack for the opposite purpose, i.e., rather than computing the perturbation to degrade the separation,  we compute an imperceptible perturbation called amicable noise to improve the separation. Experimental results show that the proposed approach selectively improves the performance of the targeted separation model by 2.23 dB on average and is robust to signal compression.
Moreover, we propose multi-model multi-purpose learning that control the effect of the perturbation on different models individually.

\end{abstract}
\section{Introduction}
\label{sec:intro}

Audio source separation has been intensively studied over the last decade \cite{OzerovF10,LiutkusFRPD14,Roux15,FitzgeraldLB16,Takahashi17} owing to its wide range of applications such as karaoke, remixing, spatial audio, and many other downstream tasks. Although the separation performance has greatly improved thanks to recent advances in deep neural network (DNN)-based methods, it remains far from perfect in many challenging scenarios including the separation of music containing many instrumental sounds mixed in a stereo format
\cite{JanssonHMBKW17,Takahashi18MMDenseLSTM,JLee2019,Liu2019mss,defossez2019music,Takahashi19, Takahashi21}.
In some cases, such as music production, sources can be assumed to be known during the mixing stage. Informed source separation (ISS) takes advantage of this and computes side-information during the so-called \textit{encoding} stage. 
Side-information can be either embedded into mixtures by using watermark approaches \cite{Parvaix10, Liutkus12} or simply transmitted along with the mixtures \cite{Ozerov11}. Separation models are designed to utilize the side-information to improve the performance. 

In this work, different from previous works, we adopt a pretrained DNN-based separation model that is trained without any side-information for ISS. Rather than modifying the pretrained separation model to use the side-information, we compute an imperceptible perturbation that is carefully designed to improve the separation of the model and add it to the mixture. 
The proposed approach is closely related to adversarial examples, which were originally  discovered in image classification \cite{Szegedy2014}, that is, imperceptibly small perturbations can significantly alter DNN predictions.
The proposed method in this paper can be seen as an application of adversarial attacks for the opposite purpose, namely, the perturbation is computed to improve the separation rather than degrade it. In this analogy, we refer to samples computed by the proposed method as \textit{amicable examples}.  However, the effectiveness of amicable examples is unclear as they can have potentially different properties from adversarial examples. This is because (i) while (untargeted) adversarial examples only need to be apart from targets and many possible perturbation can degrade the separation, amicable examples have concrete targets; thus, amicable examples may be difficult to find or less effective; (ii) if the loss curve becomes flat around the target $y$ but becomes steep away from the target, the improvement of an amicable example may be not as significant as that of an adversarial example; (iii) amicable examples may be more prone to stacking with local optima. In our experiment, we provide both quantitative and qualitative evaluations of the effect of amicable examples. 

An advantage of the proposed method is that since the separation model is not modified to use side-information, the model can be used for both standard mixtures and amicable examples in a unified manner. 
Moreover, we show that amicable examples are robust against audio signal compression, which allows us to transmit amicable examples at low bit rates.
As shown in our experiment, amicable examples can selectively improve the performance of the targeted separation model and have very limited effects on untargeted separation models. Although this is often a desirable property, having explicit control of the effects of amicable examples on multiple models is more desirable. To this end, we propose the use of multi-model multi-purpose perturbation learning (MMPL) to control the effect of perturbation in both positive and negative ways depending on the separation model.

The contributions of this work are summarized as follows:
{
\setlength{\leftmargini}{12pt} 
\begin{enumerate}
\item We propose amicable examples, the opposite optimization problem to adversarial examples, and apply them to ISS. The proposed method allows us to use the same separation model universally under both informed- and non-informed conditions.
\item We investigate the effectiveness of amicable examples on targeted and untargeted models and show the selective effectiveness for the targeted model. We further show the robustness of amicable examples against distortions caused by signal compression.
\item We further propose MMPL to control the effects of the perturbation against multiple models individually.
\item We show that, by using MMPL, amicable and adversarial examples can co-exist, namely, a perturbation can significantly improve the performance of some models and significantly degrade the performance of others.
\end{enumerate}
}

\section{Related works}
\textbf{Side-information approaches}:
In \cite{Parvaix10}, the modified discrete cosine transform (MDCT) coefficients are used to form the side-information, and separation is performed by estimating a mask from the encoded MDCT coefficients at each time-frequency point by assuming the sparsity of sources. 
In \cite{Ozerov11, Liutkus12}, local Gaussian models (LGM) are adopted to solve ISS problems. Bl\"{a}ser \etal applied non-negative tensor factorization (NTF) to ISS, where factorized matrices are compressed and transmitted as the side-information and reconstructed matrices are used to calculate the parameters of the Wiener filter \cite{Max18}.
Another line of works that closely related to ISS is to use extra information such as a music score or text \cite{Miron17,Manilow20,Kinoshita2015TextinformedSE}, or to use an extra model trained for other tasks such as automatic speech recognition \cite{Takahashi20} to leverage knowledge from other domains. 
Existing ISS approaches heavily rely on the side-information, and separation models are specifically designed to use the side-information. Therefore, such models cannot perform separation or perform poorly if the side-information is not available.
\\
\textbf{Adversarial examples}: Since adversarial examples can be a crucial problem for many DNN-based systems, they have been intensively investigated from different aspects including attack methods \cite{Su2019}, defense methods \cite{Madry2018, Bai2019}, transferability \cite{Dong2018, Wu2020}, and the cause of network vulnerabilities \cite{Goodfellow2015, Ilyas2019}. Recently Takahashi \etal investigated adversarial examples on audio source separation and reported that some attack methods are effective with limited transferability \cite{Takahashi21adv}.

\begin{figure}[t]
  \centering
  \includegraphics[width=80mm]{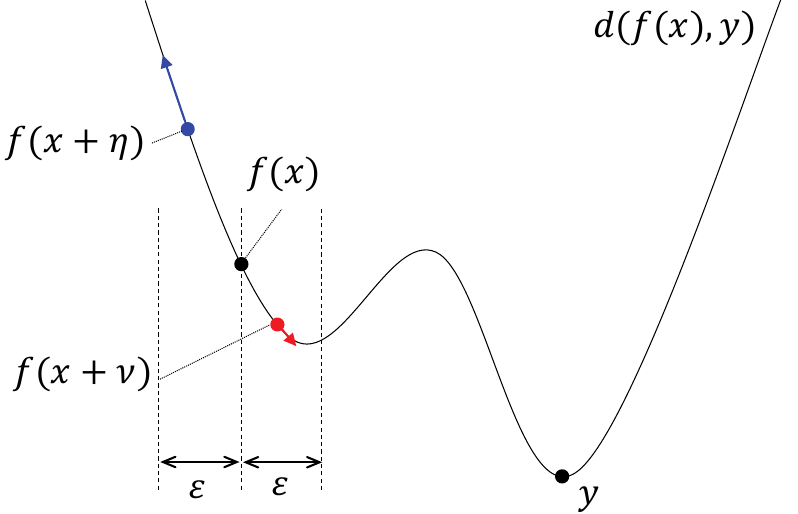}
  \caption{Simplified visualization of a loss curve for the separation $f(x)$. For simplicity, we consider the $l_\infty$ norm for the constraint ($\|\nu\|_\infty<\epsilon$). For the adversarial example $x+\eta$, the input $x$ is perturbed towards the direction to increase the criterion $d()$ within an $\epsilon$-ball  while the perturbation $\nu$ for the amicable example decreases it. The minimum loss can be obtained when $x+\nu=y$, which may be outside of the $\epsilon$-ball.}
  \label{fig:advamic}
\end{figure}

\section{Amicable-example-based informed Source Separation}
Given $N$ sources $y=[y_1, \dots, y_N]$ and a mixture $x=\sum_{i=1}^N y_i$, a DNN-based separation model $f_\theta()$ is trained to minimize the expectation of a training criterion $d()$ across data $D$ as
\begin{equation}
    \min_\theta \mathbb{E}_{(x,y)\in D}[d(f_\theta(x), y)],
    \label{eq:ss}
\end{equation}
where $\theta$ denotes network parameters. A typical choice for $d()$ is $l1$ or $l2$ distance. We use $l2$ distance in this work. Unlike conventional ISS, where the separation model is designed to use the side-information $\psi_\omega(y)$ encoded from $y$ and optimize the parameters $\theta, \omega$ as $\min_{\theta, \omega}d(f_\theta(x, \psi_\omega(y)),y)$, we fix the separation model parameters $\theta$ and instead compute a perturbation $\nu$ that minimizes the criterion under a constraint $\mathcal{C}$ on the perturbation as
\begin{equation}
    \min_{\nu \in \mathcal{V}} d(f_\theta(x+\nu), y),~ \mathcal{V} = \{ \nu ~|~ \mathcal{C}(\nu) < \epsilon\}.
    \label{eq:amic}
\end{equation}
The perceptibility of the perturbation $\nu$ highly depends on the input mixture $x$ to be added, for example, low-level noise can be perceptible when the mixture is also low-level, and high-level noise can be hardly perceptible when the mixture is also high-level. To incorporate the masking effect,  we use short-term power ratio (STPR) regularization \cite{Takahashi21adv} as the constraint, i.e.,
\begin{equation}
   \mathcal{C}_{\STPR}(\nu) = \|\vartheta(\nu, l) / \vartheta(x, l)\|_1,
    \label{eq:mask}
\end{equation}
where $\vartheta(\nu, l) = [\|\nu_1\|_2, \cdots, \|\nu_N\|_2]$ is the framewise $l2$ norm function, which computes the norms of short frames $\nu_n=[\nu(t_n),\cdots,\nu(t_n+l)]$ of length $l$ starting from time index $t_n=(n-1)l$.
We use $l$ = 4096 samples. 
Unlike adversarial examples, where the perturbation can be arbitrarily large without the constraint on the magnitude of the perturbation, amicable noise $\nu$ can be \textit{self-regularized}; the perturbation may not become too large without any constraint because injecting too large perturbation in the mixture itself makes the separation difficult and thus, the amicable perturbation may inherently need to be small to minimize the separation error. Nevertheless, we found that the constraint is essential not only for regularizing the magnitude of the perturbation but also for robustly obtaining improvements in the separation.

By introducing a Lagrange weight $\lambda$, \eqref{eq:amic} can be solved by minimizing the loss function $L$ using stochastic gradient descent: 
\begin{equation}
   L(\nu) = \|f(x+\nu)- y\|_2^2 + \lambda\mathcal{C}_{\STPR}(\nu).
    \label{eq:amicloss}
\end{equation}
We omit $\theta$ for the sake of clarity.

If the negative of the first term is used instead, \eqref{eq:amicloss} results in the loss function for the adversarial example. However, optimization behavior can be different depending on the loss surface, as shown in Fig.~\ref{fig:advamic}. If the loss curve becomes flat around the (local) optimal point but becomes steep away from the optimal point, the improvement by amicable examples may not be as significant as that by adversarial examples and vice versa.

\section{Incorporating multiple models for multiple purposes}
An amicable example perturbs the mixture towards the direction where the separator \textit{believes} it sounds more like the target sources. A natural question is ``\textit{Is the amicable example universal for other separation models?}". As shown in our experiment in Sec.~\ref{sec:untargeted}, we found that amicable example is specific to the separation model used to compute it, which we call the \textit{targeted model}, and it does not markedly improve the performance of untargeted models. This property is useful when one wants to minimize the side-effect on other models or to design a system where a targeted model exclusively benefits from the amicable example. However, it is more useful to have individual control in the separation of multiple separation models. To this end, we propose MMPL as 
\begin{equation}
   L(\nu) = \sum_i\alpha_i\|f^i(x+\nu), y\|_2^2 + \lambda\mathcal{C}_{\STPR}(\nu),
    \label{eq:mmpl}
\end{equation}
where $f^i$ denotes the $i$th separation model and $\alpha_i$ the weight to control the effect on each model. Note that $\alpha_i$ can be a negative value and in such a case, the term promotes the perturbation to be an adversarial example for model $f^i$. When all $\alpha_i$ are negative and $\sum_i\alpha_i=-1$, the loss becomes similar to the adversarial attack against an ensemble of models \cite{Liu17ensambleattack}. However, MMPL is more general and flexible as it can produce both amicable and adversarial examples at the same time depending on the model, i.e., suppose $\Gamma=\{i|\alpha_i>0\}$ and $\Lambda=\{i|\alpha_i<0\}$, the perturbation $\nu$ acts as the amicable example for models $f^{i\in\Gamma}$ but acts as adversarial example for models $f^{i\in\Lambda}$. To the best of our knowledge, this is the first attempt to learn a perturbation that serves as both amicable and adversarial examples simultaneously.

\label{sec:perturblevel}
\begin{figure}[t]
  \centering
  \includegraphics[width=\linewidth]{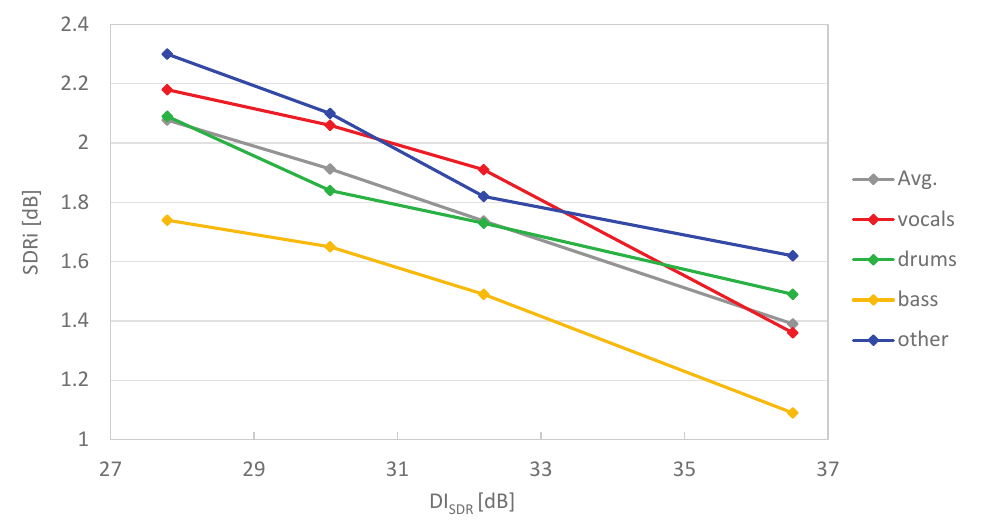}
  \caption{SDR improvement with different amicable noise levels. Higher $\DISDR$ indicates lower noise level.}
  \label{fig:perturblevel}
\end{figure}

\section{Experiments}
\label{sec:exSS}
\subsection{Setup}
Experiments are conducted on the \textit{test} set of the MUSDB18 dataset \cite{sisec2018}, which contains 50 songs recorded in stereo format at 44.1 kHz. 
Four sources ({\it vocals, drums, bass, other}) and their mixture are available for each song. To speed up the evaluation for our extensive experiments, we crop 10s clips from each song and use them for the evaluation. 

The signal-to-distortion ratio (SDR) is used as the evaluation metric of separation performance. As in SiSEC 2018 \cite{sisec2018}, SDR is computed using the {\it museval} package and the median over all tracks of the median of each track is reported. To evaluate how much the mixture is distorted by the perturbation, we use the SDR between the mixture $x$ and the perturbed mixture $x+\nu$, 
\begin{equation}
    \DISDR = \SDR(x, x+\nu), 
    \label{eq:metric}
\end{equation}
which we call the degradation of input $\DISDR$.

For the separation models, we use three open-source libraries, namely Open-Unmix (UMX) \cite{stoter19}, D3Net \cite{Takahashi21}, and Demucs \cite{defossez2019music}, to ensure a variety of separation algorithms. UMX is based on bidirectional long-short term memory (BLSTM) layers and performs the separation in the frequency domain. D3Net is a convolutional neural network and also operates in the frequency domain. Demucs consists of both convolution and BLSTM layers and performs separation in the time domain. All models are trained on the MUSDB18 \textit{train} set. In addition, UMX and Demucs have their variants: UMX$_{{\rm HQ}}$ is trained on the uncompressed MUSDB18 \textit{train} set and Demucs$_{ex}$ is trained with 150 additional songs.

The initial perturbation is a uniform noise $[-\epsilon, \epsilon], \epsilon=0.01$, and is optimized using Adam for 300 iterations with the learning rate of 0.01.

\begin{table}[t]
    \caption{\label{tab:subj} {\it Subjective test on perceptibility of amicable examples.}}
    \vspace{2mm}
    \centering{
    \begin{tabular}{c | c} 
    \hline
    $\DISDR$ [dB] & 	Accuracy\\
    \hline\hline
    27.8 &	51.0\%\\
    30.1 &	49.0\%\\
    \hline
    \end{tabular}
    }
\end{table}

\begin{table}[t]
    \caption{\label{tab:untargeted} {\it SDRs of the separation of original mixtures and amicable example computed using UMX$_{{\rm HQ}}$ (in dB). The amicable example selectively improves the performance of the targeted model.}}
    \vspace{2mm}
    \centering{
      \footnotesize
    \begin{tabular}{c | c | c c c c c} 
    \hline
    Model & input & vocals & drums & bass & other & Avg.\\
    \hline\hline
    UMX$_{{\rm HQ}}$   & \multirow{4}{*}{Original}	& 6.25	& 6.24	& 5.07	& 3.40 & 5.24\\
    Demucs  &   & 6.71	& 5.92	& \textbf{5.31}	& 2.41  & 5.09\\
    D3Net   &   & \textbf{7.08}	& \textbf{6.79}	& 5.08	& \textbf{3.56} & \textbf{5.63}\\
    UMX     &   & 6.65	& 5.91	& 4.86	& 3.39  & 5.20\\
    \hline
    UMX$_{{\rm HQ}}$   & \multirow{4}{*}{\begin{tabular}{p{0.9cm}}Amicable \\{\scriptsize (UMX$_{{\rm HQ}}$)}\end{tabular}}  & \textbf{8.44}	& \textbf{8.03}	& \textbf{6.76}	& \textbf{5.61} & \textbf{7.21}\\
    Demucs	&	& 6.66	& 5.96	& 5.49	& 2.51  & 5.16\\
    D3Net   &   & 7.18	& 6.73	& 5.12	& 3.62  & 5.66\\
    UMX     &   & 7.53	& 6.74	& 5.66	& 4.46  & 6.10\\
    \hline
    \end{tabular}
    }
\end{table}

\begin{table*}[t]
    \caption{\label{tab:mmpl} {\it SDRs for separation of perturbed samples computed using MMPL in two scenarios ($\alpha_i$ are both positive and $\alpha_i$ have opposite signs). Values in brackets indicate the SDR improvement over the separation of the original mixture.}}
    \vspace{2mm}
    \centering{
    \begin{tabular}{c | c | c | c c c c c} 
    \hline
    Model & $\alpha_{i}$ & $\DISDR$ & vocals & drums & bass & other & Avg.\\
    \hline\hline
    UMX$_{{\rm HQ}}$   & positive &\multirow{2}{*}{29.21}  & 7.90 (+1.65)	& 8.26 (d+2.02)	& 6.22 (+1.15)	& 5.08 (+1.68) & 6.87 (+1.63)\\
    Demucs$_{ex}$	& positive  &	& 9.10 (+1.60)	& 9.89 (+2.01)& 9.75 (+2.19)	&5.87 (+2.56)  &8.65 (+2.09)\\
    \hline
	UMX$_{{\rm HQ}}$   & positive  & \multirow{2}{*}{28.82}	& 7.89 (+1.64)	& 8.26 (+2.02)	& 6.22 (+1.15)	& 5.08 (+1.68)	& 6.86 (+1.62)\\
	Demucs$_{ex}$ & negative& 	& 0.51 (-6.99)	& 0.5 (-7.38)	& 1.47 (-6.09)	& -1.08 (-4.39)	& 0.35 (-6.21)\\
    \hline
    \end{tabular}
    }
\end{table*}

\subsection{Level of amicable noise and separation improvement}
First, we investigate the relationship between the level of perturbation and separation performance improvement. We use UMX$_{{\rm HQ}}$ for the evaluation and set different $\lambda$ values ($[10, 20, 40, 100]$) in \eqref{eq:amicloss} to control the perturbation level. Fig. \ref{fig:perturblevel} shows the SDR improvement $\SDRi$ over the original mixture with different $\DISDR$. As expected, the SDR improvement becomes more significant with increasing perturbation level. For 27.8 dB $\DISDR$, an improvement of more than 2 dB is obtained on average.
To evaluate the perceptibility of the amicable noise, we conduct a subjective test similarly to the double-blind triple-stimulus with hidden reference format (ITU-R BS.1116), where the reference is the original mixture and either A or B is the same as the reference and the other is an amicable example. Evaluators are asked to identify which one of A or B is the same as the reference. We test two amicable noise levels and 40 audio engineers evaluated five songs of 10~s duration for each noise level. Table \ref{tab:subj} shows that the accuracy of correctly identifying the reference is close to the chance rate (50\%) at $\DISDR$ of 27.8 dB; thus, the amicable noise is imperceptible.

\subsection{Effects on untargeted models}
\label{sec:untargeted}
Next, we test the amicable example on untargeted models. The amicable example is computed using UMX$_{{\rm HQ}}$ and tested on different separation models. Table \ref{tab:untargeted} shows the SDR values computed on the separations of the original mixture and amicable examples. By comparing the results of the original mixture and amicable example for each model, we observe that the amicable example significantly improves the SDRs of the targeted model UMX$_{{\rm HQ}}$ but only slightly improves the SDRs of Demucs and D3Net. This indicates that the loss surfaces \eqref{eq:amic} of these models are very different and thus the amicable noise is not generalized to different models. In contrast, the SDR improvement of UMX is more significant than that of Demucs and D3Net, probably because the architectures of UMX and UMX$_{{\rm HQ}}$ are identical and they were trained on very similar datasets (only their high-frequency components are different); thus, their loss surfaces are probably also similar.

\subsection{Robustness against signal compression}
\label{sec:robustness}
\begin{figure}[t]
  \centering
  \includegraphics[width=\linewidth]{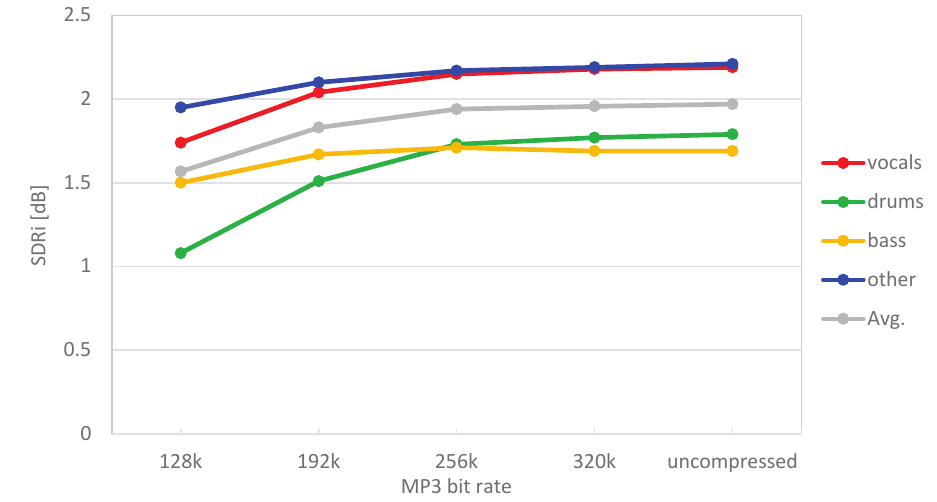}
  \caption{SDR improvement with different MP3 compression bit rates.}
  \label{fig:compress}
\end{figure}

As audio signals are often compressed to reduce the bandwidth or file size for transmission, it is important to assess the robustness of an amicable example against compression to verify its usability in realistic scenarios. To this end, we study how SDR improvements change by compressing the amicable example using an MP3 encoder with different compression levels. Fig. \ref{fig:compress} shows that even after the amicable example is compressed with 256 kbps, the SDR improvement over the original mixture is nearly the same as that of the uncompressed example. Although more aggressive compression rates slightly degrade the effectiveness, the amicable example still improvse the SDR by 1.57 dB on average at 128 kbps.

\subsection{Amicable adversarial example with MMPL}
\label{sec:MMPL}

Finally, we evaluate MMPL in two scenarios using UMX$_{{\rm HQ}}$ and Demucs$_{ex}$. In the first scenario, $\alpha_{i}$ in \eqref{eq:mmpl} is set to be positive for both UMX$_{{\rm HQ}}$ and Demucs$_{ex}$. In this case, the perturbation is computed to improve both models. In the second scenario, we use a negative $\alpha_i$ for Demucs$_{ex}$ instead. This makes the perturbation amicable for UMX$_{{\rm HQ}}$ but adversarial for Demucs$_{ex}$. To balance the magnitude of loss for both models, we set $(\alpha_{\rm{UMX}}, \alpha_{\rm{Demucs}})$ to be $[1,100]$ for the former case and $[1,-100]$ for latter.
The results are shown in Table \ref{tab:mmpl}. As observed, when we include both models to compute the amicable example, the performance of both models are improved, which is not the case when only one model is used, as shown in Sec. \ref{sec:untargeted}. More interestingly, in the second scenario, where we use opposite signs for the two models, we observe that the same perturbation significantly improves UMX$_{{\rm HQ}}$ but significantly degrades Demucs$_{ex}$. The results show that we can design a perturbation to be both an amicable example and an adversarial example depending on the model. We believe that this finding is important as it is closely related to security applications, e.g., even if a sample can be separated well with some models, it still can be an adversarial example for other models. We will further investigate this in the future.

\section{Conclusion}
We propose amicable example-based informed source separation, where an imperceptible perturbation from the mixture is computed to improve the separation. Experimental results show that amicable examples selectively improve the performance of the targeted model and are robust against the audio compression. We further propose multi-model multi-purpose learning (MMPL) to individually control the effect of the perturbation for multiple models. MMPL is shown to be capable of computing a perturbation that works as both an amicable example and an adversarial example depending on the model.

\ninept

\bibliographystyle{IEEEbib}
\bibliography{MIR,bss,other,advexamp}


\end{document}